\documentclass{article}

\usepackage{arxiv}

\usepackage[utf8]{inputenc} % allow utf-8 input
\usepackage[T1]{fontenc}    % use 8-bit T1 fonts
\usepackage{hyperref}       % hyperlinks
\usepackage{url}            % simple URL typesetting
\usepackage{booktabs}       % professional-quality tables
\usepackage{amsfonts}       % blackboard math symbols
\usepackage{nicefrac}       % compact symbols for 1/2, etc.
\usepackage{microtype}      % microtypography
\usepackage{lipsum}
\usepackage{graphicx}
\graphicspath{ {./images/} }

\title{Evaluating Language Models for Threat Detection in IoT Security Logs}        
\author{
Jorge J. Tejero-Fern\'andez \\
GMV\\
C. de Isaac Newton, 11\\
Tres Cantos, Madrid, Spain\\,
\And
Alfonso S\'anchez-Maci\'an \\
Universidad Carlos III de Madrid\\
Avda de la Universidad 30\\
Legan\'es, Madrid, Spain\\
\texttt{alfonsan@it.uc3m.es},
}

\begin{document}
\maketitle
\begin{abstract}
Log analysis is a relevant research field in cybersecurity as they can provide a source of information for the detection of threats to networks and systems. This paper presents a pipeline to use fine-tuned Large Language Models (LLMs) for anomaly detection and mitigation recommendation using IoT security logs. Utilizing classical machine learning classifiers as a baseline, three open-source LLMs are compared for binary and multiclass anomaly detection, with three strategies: zero-shot, few-shot prompting and fine-tuning using an IoT dataset. LLMs give better results on multi-class attack classification than the corresponding baseline models. By mapping detected threats to MITRE CAPEC, defining a set of IoT-specific mitigation actions, and fine-tuning the models with those actions, the models are able to provide a combined detection and recommendation guidance.
\end{abstract}

\keywords{Large language models, Internet of Things, Logs, Threat detection, fine-tuning}

\section{Introduction}
\label{sec1}
Anomaly detection has long been a fundamental technique for identifying threats in cybersecurity. Initially, threat detection relied on static rules or expert-defined signatures, which were a limitation because it only detected previously known events. Over time, as attacks became more sophisticated and environments more variable, heuristic techniques, specifically those based on Machine Learning (ML), were adopted. These methods can identify both known threats without explicit rules and new threats that have not yet been observed. Some of the best-known approaches include Principal Component Analysis (PCA) \cite{xu_detecting_2009}, Isolation Forest \cite{liu_isolation_2008}, and Support Vector Machine (SVM) \cite{wang_anomaly_2004}, all of which were successfully applied in environments with structured, balanced, and stable data.

However, these methods exhibit significant limitations when applied to unstructured, sequential, and evolving data streams like system logs. 

Log analysis is a highly relevant research field in cybersecurity. Logs, especially in the context of the IoT, are generated in massive volumes by heterogeneous devices and frequently changing software versions, posing challenges for parsing and feature extraction. 

Logs remain a cornerstone of modern computing systems. They are critical for system monitoring, debugging, and security auditing, since they record system state and offer valuable insights into performance and potential issues, including security incidents \cite{han_loggpt_2023}. Consequently, they are a key information source in Security Operation Centers (SOCs), where log analysis enables alert prioritization, threat detection, and reduction of manual workload. 

However, logs pose major challenges due to their variability, high dimensionality, large volume, and complex structure \cite{han_loggpt_2023, cui_logeval_2024}. Recent studies have introduced the concept of log drift or log instability, which describes how previously trained detection models lose accuracy when log formats or semantic structures change \cite{akhtar_llm-based_2025}. This issue is especially critical in IoT, where devices may be frequently updated or reconfigured, producing new log patterns that were not present during initial training.

These characteristics make traditional approaches infeasible and have driven investigation into more adaptive solutions. Among these is the use of LLMs to interpret event sequences without requiring transformation, providing a richer understanding of context (e.g. \cite{ferrag_revolutionizing_2024}). Effective Artificial Intelligence (AI) based threat detection requires the availability of heterogeneous datasets to train the models. Fortunately, a number of them, including labeled IoT logs, have been provided in different studies (e.g. \cite{toniot}, \cite{botiot}).

The main contributions of this work are as follows:
\begin{enumerate}
    \item Traditional ML models are compared to up-to-date open source LLMs for the detection of threats in IoT security logs extracted from a dataset containing both normal and malware traffic. The comparison is performed for binary (threat vs. no threat) and multiclass classification (type of attack).
    \item ML model training  and LLM fine-tuning are performed on a small subset of data from the same dataset to check the performance under data-scarce conditions.
    \item For LLMs, fine-tuning is compared against zero-shot and few-shot prompting.
    \item A method is proposed to include mitigation recommendations into the responses generated by the fine-tuned LLM.
\end{enumerate}

\section{Background}
\label{background}

Before the advent of Transformer-based architectures, various machine learning approaches were evaluated for anomaly detection in logs. Classical models such as PCA, Isolation Forest, and SVM, though effective on simple datasets, exhibited significant limitations with long sequences, non-standard structures, or shifts in log patterns. For instance, PCA achieved an F1-score of only 0.08 on the HDFS dataset \cite{han_loggpt_2023}. Similarly, SVM’s performance \cite{tian_cldtlog_2023} fell to an F1-score of 0.46 compared to 0.9999 for the CLDTLog model on the BGL dataset, which used BERT. The CLDTLog training and prediction architecture employs BERT trained with a combination of Triplet Loss and Focal Loss to detect anomalies without parsing. This is why classical models failed to capture temporal relationships or contextual patterns, which limits their applicability to sequential, unstructured logs.

These limitations led to the emergence of models capable of representing sequential and semantic relationships, such as LogBERT\cite{guo_logbert_2021}. Proposed in 2021 , LogBERT was one of the first models to apply BERT-style architectures to log analysis. The model employs a bidirectional encoder trained via two self-supervised tasks: MLKP (Masked Log Key Prediction) and VHM (Volume of Hypersphere Minimization). The first enables the model to learn semantic dependencies within log sequences without requiring labels, while the second forces representations of normal sequences to cluster in the embedding space, thereby facilitating the detection of anomalous deviations.

LogBERT managed to outperform models such as DeepLog \cite{du_deeplog_2017} and LogAnomaly \cite{zhou_loganomaly_2019}, especially on large datasets like BGL and Thunderbird, which are widely used in the literature \cite{oliner_what_2007}. This has made it a common reference in subsequent works such as LogGPT\cite{han_loggpt_2023} or CLDTLog\cite{tian_cldtlog_2023}, which extend its approach toward more advanced generative or multitask architectures.

However, as previously pointed out, logs present limitations and one of the greatest challenges in log analysis is structural variability over time. Logs constantly evolve due to software updates, configuration changes, or system upgrades, causing what is known as log instability or log drift. This phenomenon means that models trained on one log format often fail or generate many false positives when formats or semantic structures change.

This problem has been widely studied. Hadadi et al.\cite{hadadi_llm_2025} introduced the ADUL framework (Anomaly Detection on Unstable Logs), evaluating both classical models and LLMs against various log modifications such as event deletion, duplication, or reordering. Their results show that fine-tuned GPT-3 models maintain strong performance, while traditional techniques and LLMs in zero-shot settings suffer drastically under minor structural changes.

The LOGEVOL-Hadoop \cite{cui_logeval_2024} benchmark assesses the impact of real software version changes (Hadoop~2 vs. Hadoop~3), demonstrating that models trained on one version do not generalize well to another. Cui et al. in their LogEval study \cite{cui_logeval_2024} highlight that this lack of generalization is a major weakness, especially when parsing fails or the vocabulary changes. They also note that many LLMs hallucinate when logs differ significantly from training data.

A systematic review \cite{akhtar_llm-based_2025} identified log instability as a key unresolved challenge in applying LLMs to logs. They emphasize that fine-tuned models often require frequent retraining, which is impractical in dynamic IoT environments. 

This structural drift has spurred the development of more robust, adaptive methods capable of reasoning about unseen inputs, vital for IoT systems’ heterogeneous, rapidly changing landscapes.

A recent study by Balan et al.~\cite{balanfine} reinforces the growing interest in the use of fine-tuned LLMs for security tasks. The authors compare LLaMA 2, Mistral, and Mixtral against 18 traditional machine learning algorithms for malware detection based on API call logs. Their results show that fine-tuned LLMs achieve higher generalization and recall. However, they also point out significant drawbacks, including high inference times and memory usage, which hinder real-time deployment. Instead, the authors suggest these models are more suitable for advanced analysis layers such as sandboxes or EDR/XDR platforms.

These findings highlight both the promise and the practical trade-offs of deploying LLM-based detection systems, especially in resource-constrained environments like IoT.

\subsection{Datasets in IoT Environments}

Developing and evaluating anomaly detection models for logs requires datasets that reflect real-world complexity, variability, and characteristics. In IoT, where devices generate massive data volumes, datasets must include benign and malicious traffic as well as diverse attack types and structures suitable for automated analysis by language models.

Key publicly available datasets include TON-IoT\cite{toniot}, which focuses on network traffic and telemetry from simulated smart home and smart city environments; though versatile, it centers on sensor data, limiting its utility for textual log analysis. Bot-IoT\cite{botiot} is widely used for botnet and denial-of-service attack detection but suffers from significant class imbalance.

IoT-23 \cite{iot23} provides real and simulated network flows from IoT devices but lacks multiclass labels, complicating complex classification schemes. 

The CIC IoT datasets\cite{cicdatasets} cover various attack scenarios but are fragmented and ununified across subsets. 

In industrial settings, RT-IoT~2022~\cite{rtiot2022} offers high-quality data from real-time devices and networks but with limited volume and attack coverage.

Finally, the Edge-IIoTset dataset~\cite{edgeiiot} was specifically designed to realistically represent IoT and IIoT scenarios. It contains over 20 million labeled samples generated via a multilayer infrastructure, encompassing diverse attacks, real sensors, and industrial protocols, offering a robust foundation for both traditional machine learning and large language model approaches.

\subsection{LLMs for Log-Based Threat Detection}

Early explorations for using language models for log-based threat detection consisted in applying prompt engineering without retraining, using zero-shot and few-shot learning. Hadadi et al. evaluated GPT-4 on unstable logs\cite{hadadi_llm_2025}, achieving a zero-shot F1-score of 0.45 and improving to 0.93 in few-shot settings. This work exposed zero-shot limitations when logs contain unseen structures.

In a similar approach, the LogEval benchmark \cite{cui_logeval_2024} assessed 18 LLMs on tasks like parsing, anomaly detection, fault diagnosis, and log summarization. Few-shot methods improved parsing and diagnosis ($F1 > 0.9$) but introduced overfitting and repetition in anomaly detection; zero-shot remained more stable for that task.

Guastalla et al. applied few-shot prompting to detect DDoS attacks in IoT environments, achieving 0.7 accuracy without fine-tuning \cite{guastalla_application_2024}. While promising, results depend heavily on prompt design and context quality.

As prompting alone was insufficient for reliable detection, especially on unstable or imbalanced logs, research shifted to fine-tuning LLMs with domain-specific labeled examples, boosting accuracy without manual prompt engineering.

HackMentor \cite{zhang_hackmentor_2023} applied Low-Rank Adaptation (LoRA)\cite{hu2021loralowrankadaptationlarge} to fine-tune LLaMA and Vicuna models for cybersecurity tasks. HackMentor models achieved 10--25\% improvements over base models, matching ChatGPT performance while enabling offline deployment, lower cost, and better data control.

SecurityBERT \cite{ferrag_revolutionizing_2024}, a lightweight BERT-based model optimized for CPU-only IoT/IIoT environments, achieved 98.2\% accuracy on the Edge-IIoTset dataset, outperforming more complex alternatives and demonstrating feasibility for IoT deployment.

LLM4ITD \cite{zhang_llm4itd_2024} focused on insider threat detection in user logs, combining custom prompt generation from user behavior with LoRA fine-tuning. LLaMA-7B models reached AUCs above 0.99. 

Another work proposed Cyber LLMs \cite{batmaz_building_2024}, foundational models trained from scratch only on cybersecurity logs. These JSON-structured models outperformed generalist LLMs in threat simulation, synthetic log generation, and anomaly detection in real SOC pipelines, showing that well-trained smaller architectures can compete at lower cost.

Beyond prompting and fine-tuning, several works introduced advanced methods for anomaly detection, particularly under limited supervision, class imbalance, or high structural variability.

CLDTLog \cite{tian_cldtlog_2023} combines contrastive learning (Triplet Loss) with Focal Loss in a BERT setup, improving representation learning and handling imbalanced classes. It achieved F1-scores of 0.9971 on HDFS and 0.9999 on BGL, vastly outperforming comparative methods. Even with only 1\% of BGL’s training data, it reached an F1 of 0.9993, showing strong generalization.

LogGPT \cite{han_loggpt_2023} uses reinforcement learning (PPO) to optimize Top-K reward metrics, aligning training with anomaly detection goals. It achieved F1 $>$ 0.98 on Thunderbird and ~0.90 on HDFS.

The LLM4Sec pipeline \cite{karlsen_benchmarking_2024} fine-tunes various LLMs (BERT, RoBERTa, GPT-2, GPT-Neo) for binary log classification, finding DistilRoBERTa achieves an average F1-score of 0.998 across six datasets.

Parserless architectures, such as Ithy Labs’ model\cite{labs_building_2025}, process logs without predefined templates, using LLMs as flexible parsers and Pydantic for structural validation—streamlining real-time log ingestion and reliable alert generation.

MSIVD \cite{yang_security_2024} (Multitask Self-Instructed Vulnerability Detection) integrates vulnerability detection, textual explanation, and error localization via dialogue-style training and a lightweight Graph Neural Network for data flow analysis.

Finally, RedChronos \cite{li_redchronos_2025} integrates multiple LLMs in real SOCs, combining predictions via weighted voting and semantically guided prompt evolution. It reduced manual analysis by 90\% and achieved recall up to 0.99 on CERT 5.2 benchmarks. 

\subsection{LLMs for Mitigation Generation}

While anomaly detection is crucial, effective defense also requires contextual response recommendations. Most reviewed works focus solely on detection, leaving response automation underexplored.

LogPr\'ecis\cite{boffa_logprecis_2024} uses BERT and CodeBERT to analyze honeypot shell sessions and classify MITRE ATT\&CK TTPs with an F1-score of 0.85. However, detecting a tactic does not guarantee understanding the intention of the attacker or generating corrective actions. E.g. two persistence actions may require  different mitigations depending on the technical context.

O’Brien \cite{obrien_mitre_2024} evaluated ChatGPT, Claude 3, and LLaMA 3.1 on CTI phrase tagging through zero-shot and few-shot prompting, finding perfect recall but only around 0.5 precision, indicating high coverage but significant ambiguity.

Zhang et al. \cite{zhang_when_2025} confirmed that, despite advances in classification, vulnerability detection, and CTI generation, incident response automation remains an open challenge due to the lack of systems able to provide appropriate mitigations.

The CAPEC framework \cite{mitre_capec} offers a catalog of attack patterns and associated countermeasures, but recommendations are generic and not adapted to resource-constrained, interoperability-limited IoT contexts.

\section{Methodology}
\label{proposal}
In this section, the methodology followed in this work is described, detailing the dataset used, the data collection and preprocessing process, the implemented architectures, and the mitigation generation task.

\begin{figure}[ht]
    \centering
    \includegraphics[width=8cm]{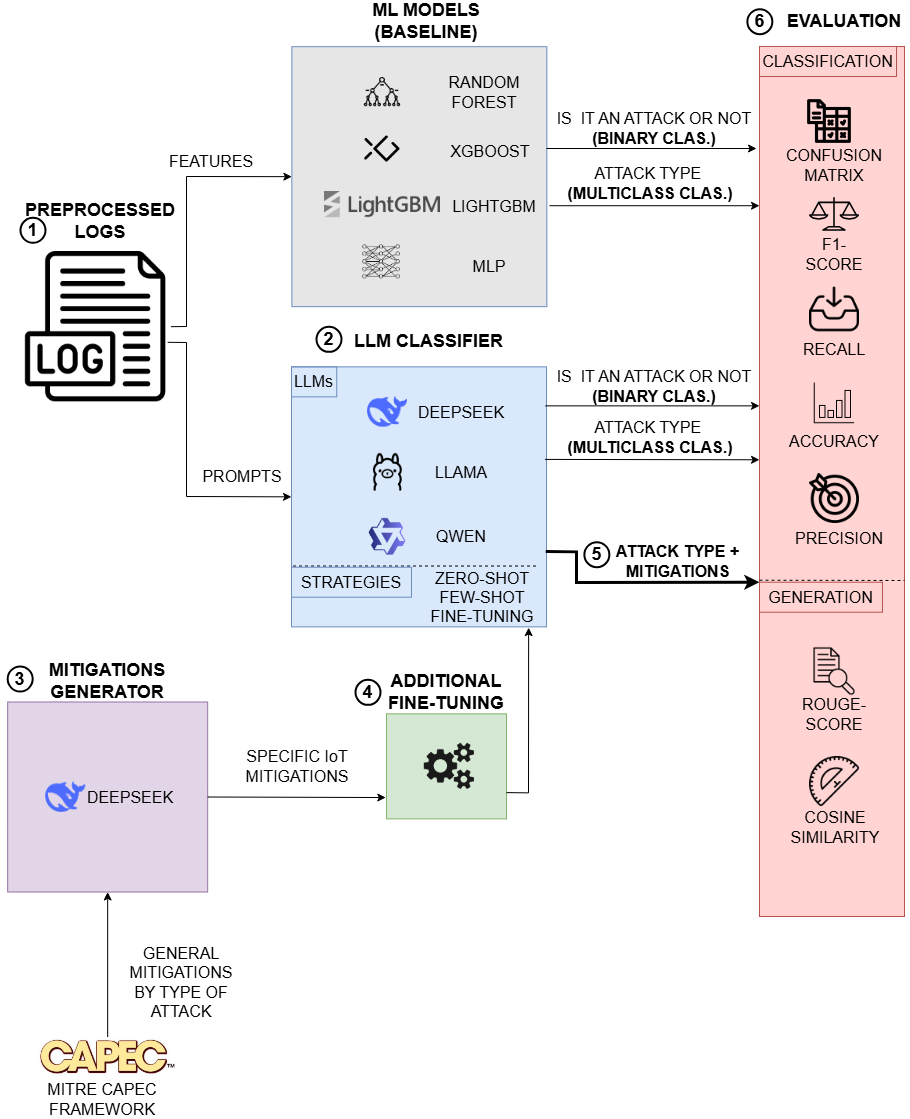}
    \caption{Methodology to evaluate the potential of ML and LLM for threat detection}
    \label{fig:methodology}
\end{figure}

The proposed methodology is designed to evaluate the potential of LLMs by comparing their performance against ML models, using a workflow organized into different consecutive phases. Figure \ref{fig:methodology} describes the methodology used, where the following steps are followed after selecting the dataset:
\begin{enumerate}
    \item \textbf{Dataset Preprocessing}: cleaning, feature selection, encoding, and normalization.
    \item \textbf{LLM Classifier}: LLM models are used by applying zero-shot, few-shot, and fine-tuning strategies, visualizing how the latter improves results. Additionally, ML models are set up as a comparative baseline.
    \item \textbf{Mitigation Generator}: Receives the attack type labels from the dataset together with the associated mitigations from the Mitre CAPEC framework, and generates a specific set of mitigations for IoT.
    \item \textbf{Additional Fine-tuning}: The specific mitigations proposed for IoT are passed to the LLMs for fine-tuning, so that for each type of detected attack, the model can generate a series of associated mitigations.
    \item \textbf{Attack classification with mitigation generation}: The LLM models identify the type of attack present in the logs and subsequently generate tailored mitigation strategies for each attack, specifically adapted to IoT environments. This integrated process of classification and response generation constitutes the core contribution of this work.
    \item \textbf{Evaluation}: With two parts, 1) evaluation of the generic LLM and ML models through classification metrics (Confusion Matrix, Accuracy, Precision, Recall, and F1-score), and 2) calculating quality metrics for the generation of mitigations (Cosine Similarity, ROUGE).
\end{enumerate}

\subsection{Dataset selection and preprocessing}
Among the different options, the Edge-IIoTset dataset was selected for its alignment with the objectives of this work. This dataset was developed in a testing environment distributed across seven layers (cloud, edge, fog, blockchain, SDN, NFV, and IoT/IIoT perception), using real devices and simulating both benign and malicious traffic using specific tools for each type of attack.

Furthermore, the dataset was processed by its authors with different tools, extracting more than 60 features that combine network information, logs, protocol traffic and system metadata. Available in CSV and PCAP formats and containing structured textual records, it is particularly well-suited for use in language models without the need for complex transformation processes. 

Table \ref{tab:edgeiiotset_distribution} shows the dataset distribution, which includes fourteen well-defined attack classes, with labels for both binary ($attack\_label$) and multi-class ($attack\_type$) classification. The main disadvantage is that some classes are imbalanced.

%\vspace{0.3cm} 
\begin{table}[t]
\centering
\caption{Class distribution in the Edge-IIoTset dataset}
\begin{tabular}{|l|r|}
\hline
\textbf{Class} & \textbf{Number of records} \\
\hline
Normal & 11,223,940 \\
\hline
Backdoor attack & 24,862 \\
DDoS\_HTTP attack & 229,022 \\
DDoS\_ICMP attack & 2,914,354 \\
DDoS\_TCP attack & 2,020,120 \\
DDoS\_UDP attack & 3,201,626 \\
Fingerprinting attack & 1,001 \\
MITM attack & 1,229 \\
Password attack & 1,053,385 \\
Port Scanning attack & 22,564 \\
Ransomware attack & 10,925 \\
SQL Injection attack & 51,203 \\
Uploading attack & 37,634 \\
Vulnerability Scanner attack & 145,869 \\
XSS attack & 15,915 \\
\hline
\textbf{Total} & \textbf{20,952,648} \\
\hline
\end{tabular}
\label{tab:edgeiiotset_distribution}
\end{table}

The data preprocessing was designed with the goal of preparing the dataset for use in both traditional machine learning models and for direct integration into language model architectures.

The following stages were applied:

\begin{enumerate}
    \item Data cleaning: removal of records with invalid values, removal of duplicates, and discarding irrelevant fields.
    \item Feature selection: using a random forest classifier, the relative importance of each attribute was evaluated. In Table \ref{tab:features_randomforest}, the seven most representative variables are selected, along with their description and importance. They were used as input for both the binary and the multiclass classification models.
    \item Encoding and normalization: Categorical variables were transformed using one-hot encoding, and numerical variables were normalized using StandardScaler, with zero mean and unit variance.
    \item Dataset split: The dataset was divided for model training. 
\end{enumerate}
\begin{table}[t]
\caption{Most relevant features selected using Random Forest}
\center
\begin{tabular}{|p{3cm}|p{1cm}|p{6cm}|p{1cm}|}
\hline
\textbf{Feature} & \textbf{Prot.} & \textbf{Description} & \textbf{Imp.} \\
\hline
\texttt{dns.qry.name.len} & DNS & Length of the queried domain name & 0.2058 \\
\hline
\texttt{mqtt.protoname} & MQTT & Name of the protocol used & 0.1388 \\
\hline
\texttt{mqtt.msg} & MQTT &Raw message transmitted & 0.1097 \\
\hline
\texttt{mqtt.topic} & MQTT & Topic of the published/ subscribed message & 0.1014 \\
\hline
\texttt{mqtt.conack.flags} & MQTT & Connection acknowledgment flags & 0.0956 \\
\hline
\texttt{tcp.options} & TCP & TCP header options & 0.0817 \\
\hline
\texttt{tcp.dstport} & TCP & Destination port of the packet & 0.0557 \\
\hline
\end{tabular}
\label{tab:features_randomforest}
\end{table}

\subsection{LLM classifier}
Three widely used and accessible open-source LLMs were selected, available through the Hugging Face platform \cite{huggingface2023}. All of them were loaded and adapted using the Unsloth library ~\cite{unsloth2024}, which allows efficient fine-tuning in resource-constrained environments. The selected models were:
\begin{itemize}
    \item DeepSeek-R1~\cite{deepseek2024}: A model based on the LLaMA architecture, which has recently gained attention for its computational efficiency.

    \item LLaMA 3.2~\cite{llama32024}: A lightweight model developed by Meta AI, used to evaluate the efficiency and accuracy of smaller models in specific tasks.

    \item Qwen 2.5~\cite{qwen2024}: A model by Alibaba, with a good balance between size and performance, particularly robust in classification and reasoning tasks.
\end{itemize}

Each model was evaluated under three usage strategies:
\begin{enumerate}
    \item Zero-shot: direct inference without additional training or examples.
    \item Few-shot: incorporation of representative examples within the prompt to guide the prediction through contextual learning.
    \item Fine-tuning: supervised training on a subset of the dataset in instruct format, allowing the model to be adapted for precise classification of the attack type.
\end{enumerate}

As a comparative baseline, several classical machine learning algorithms were trained: Random Forest, XGBoost, MLP, and LightGBM. All of them used as input the seven most representative features previously selected using Random Forest. These models were evaluated in binary classification tasks and multiclass classification tasks (identifying the specific type of attack), thus allowing the comparison with the performance of the proposed LLM models.

\subsection{Mitigation generation}
In addition to the classification task, a module was implemented for generating specific mitigations by attack type, using the Deepseek model. This choice was made because it was the model that achieved the best results, as will be shown later.

For the generation of mitigations, the MITRE CAPEC framework ~\cite{mitre_capec} was used, which provides general descriptions and countermeasures associated with widely recognized attack patterns. Each attack label in the dataset was manually mapped to its corresponding CAPEC pattern, as shown in Table \ref{tab:edgeiiotset_capec_mapping}.

\begin{table}[ht]
\centering
\caption{Mapping of Edge-IIoTset dataset attacks with CAPEC patterns}
\begin{tabular}{|p{4cm}|p{2cm}|p{5cm}|}
\hline
\textbf{Attack label (Edge-IIoTset)} & \textbf{CAPEC ID} & \textbf{CAPEC pattern name} \\
\hline
DDoS\_UDP & CAPEC-486 & UDP Flood \\
\hline
DDoS\_ICMP & CAPEC-487 & ICMP Flood \\
\hline
SQL\_injection & CAPEC-66 & SQL Injection \\
\hline
DDoS\_TCP & CAPEC-482 & TCP Flood \\
\hline
Vulnerability \_scanner & CAPEC-310 & Scanning for Vulnerable Software \\
\hline
Password & CAPEC-49 & Password Brute Force \\
\hline
DDoS\_HTTP & CAPEC-488 & HTTP Flood \\
\hline
Uploading & CAPEC-242 & Code Injection \\
\hline
Backdoor & CAPEC-523 & Malicious Software Implanted \\
\hline
Port\_Scanning & CAPEC-300 & Port Scanning \\
\hline
XSS & CAPEC-63 & Cross Site Scripting (XSS) \\
\hline
Ransomware & CAPEC-542 & Targeted Malware \\
\hline
Fingerprinting & CAPEC-224 & Fingerprinting \\
\hline
MITM & CAPEC-94 & Man-In-The-Middle Attack \\
\hline
\end{tabular}
\label{tab:edgeiiotset_capec_mapping}
\end{table}

Based on these general definitions, adaptive prompts were designed to transform generic mitigations into recommendations specific to the IoT environment.

These recommendations were then used for additional fine-tuning, with the aim of improving the coherence, precision, and practical usefulness of the mitigation-generating model. Thus with this configuration the fine-tuned model will not only provide a classification for the attack, but also a set of recommendations.

\section{Experiments}
\label{experiments}
This section introduces the environment used for the experiments, the implementation and configuration parameters.

\subsection{Infrastructure, data processing and metrics}
The infrastructure used for the project consisted in a workstation with a 8GB NVIDIA RTX 4060 GPU, with the code implemented in Python language and using different open source AI libraries. The code for the project is available at Github\footnote{\url{https://github.com/Jorge-Tejero-Fdez/ThreatLogLLM/}}. Due to limitations derived from the GPU, the experiments were executed with a subset of entries from the dataset. Additionally, Random Forest was used to reduce dimensionality and enable integration with lightweight models and LLMs without compromising performance.

The dataset was divided into three subsets: training, validation, and testing, assigning 70\% of the samples to the training set and 15\% to each of the validation and testing sets. Additionally, for the multiclass classification task, two configurations were used: one with a larger number of samples and another reduced version, with the goal of comparing model performance under different dataset sizes. The full distribution of samples for each task is shown in Table~\ref{tab:data_splits}.

%\vspace*{-\baselineskip}
\begin{table}[ht]
\centering
\caption{Number of samples in the training, validation, and test sets.}
\label{tab:data_splits}
\begin{tabular}{|p{5cm}|p{2cm}|p{2cm}|p{2cm}|p{1cm}|}
\hline
\textbf{Task} & \textbf{Training} & \textbf{Validation} & \textbf{Test} & \textbf{Total} \\
\hline
Binary classification & 8,400 (70\%) & 1,800 (15\%) & 1,800 (15\%) & 12,000 \\
\hline
Multiclass classification (full) & 10,500 (70\%) & 2,250 (15\%) & 2,250 (15\%) & 15,000 \\
\hline
Multiclass classification (reduced) & 5,250 (70\%) & 1,125 (15\%) & 1,125 (15\%) & 7,500 \\
\hline
\end{tabular}
\end{table}

The metrics used to compare the LLM/ML classification are accuracy, precision, recall, f1-score and the confusion matrix. 

The evaluation of the generated mitigations cannot be performed using the classification metrics explained earlier, as it involves text generation, and small errors such as the omission or alteration of words would drastically affect the results. For this reason, specific metrics aimed at evaluating semantic similarity and textual overlap have been employed, which are described below:
\begin{itemize}
    \item \textbf{Cosine Similarity} between the embeddings of the generated mitigations and the mitigations expected by the pre-trained model. A value close to 1 implies a high semantic similarity between the generated mitigation and the expected reference.

    \item \textbf{ROUGE-L} measures the overlap of word sequences between the generated mitigation and the expected one, based on the Longest Common Subsequence (LCS). A high ROUGE-L value indicates that the generated mitigation maintains a high degree of textual match with the expected mitigation.
\end{itemize}

\subsection{Binary and multi-class classification}
As previously pointed out, classic ML models were used as baseline references to compare the results obtained from LLM models. The classic models used in this phase were Random Forest, XGBoost, MLP, and LightGBM. All of them were used with their basic hyperparameters, detailed in Table \ref{tab:ml_hyperparameters}, as their main goal was to serve as a comparative baseline. The samples used as input included the seven previously selected features, after being normalized and encoded using standard preprocessing techniques.

\begin{table}[ht]
\centering
\caption{Hyperparameters used in the ML classification models.}
\label{tab:ml_hyperparameters}
\begin{tabular}{|p{3cm}|p{12cm}|}
\hline
\textbf{Model} & \textbf{Hyperparameters} \\ \hline
\textbf{Random Forest} & \texttt{n\_estimators} = 200, \texttt{max\_depth} = none \\ \hline
\textbf{XGBoost} & \texttt{n\_estimators} = 100, \texttt{max\_depth} = 6, \texttt{learning\_rate} = 0.1, \texttt{subsample} = 0.8, \texttt{colsample\_bytree} = 0.8, \texttt{random\_state} = 42, \texttt{eval\_metric} = \texttt{logloss} (binary) / \texttt{mlogloss} (multi-class) \\ \hline
\textbf{LightGBM} & \texttt{n\_estimators} = 300, \texttt{max\_depth} = -1, \texttt{learning\_rate} = 0.05, \texttt{num\_leaves} = 50, \texttt{min\_data\_in\_leaf} = 10, \texttt{subsample} = 0.7, \texttt{colsample\_bytree} = 0.7, \texttt{class\_weight} = \texttt{balanced}, \texttt{random\_state} = 42 \\ \hline
\textbf{MLP} & Architecture with 3 dense layers (\texttt{input → 256 → 128 → 64 → output}), activation function \texttt{ReLU}, \texttt{BatchNorm1d} after each hidden layer, \texttt{Dropout} = 0.3, optimizer \texttt{Adam}, \texttt{learning\_rate} = 0.001, \texttt{batch\_size} = 64, number of \textit{epochs} = 100 \\ \hline
\end{tabular}
\end{table}

Three large language models (LLMs) were evaluated using the Unsloth library~\cite{unsloth2024} and the Hugging Face platform~\cite{huggingface2023}. The models used were:
\begin{itemize}
    \item DeepSeek-R1-Distill-Llama-8B\footnote{\url{https://huggingface.co/unsloth/DeepSeek-R1-Distill-Llama-8B}} (unsloth/DeepSeek-R1-Distill-Llama-8B)
    \item Qwen 2.5 7B Instruct\footnote{\url{https://huggingface.co/unsloth/Qwen2.5-7B-Instruct-bnb-4bit}} (unsloth/Qwen2.5-7B-Instruct-bnb-4bit)
    \item LLaMA 3.2 3B\footnote{\url{https://huggingface.co/unsloth/Llama-3.2-3B-unsloth-bnb-4bit}}  (unsloth/Llama-3.2-3B-unsloth-bnb-4bit)
\end{itemize}
Two classification tasks were evaluated: a binary task (detection of malicious traffic versus benign) and a multi-class task (specific identification of the type of attack among the 14 available classes).

For the LLMs, three different strategies were applied for each classification task: zero-shot, few-shot, and fine-tuning.

Specific prompts were designed for binary and multi-class classification for each of the three strategies mentioned earlier, using the values of the seven selected features, structured as text in natural language. In the case of fine-tuning, a dataset in instruct format was constructed, where each sample included an instruction based on the traffic log and a binary response as the label.

In order to compare the real performance of each LLM model fairly, the same hyperparameter configuration was used in each of the three strategies (zero-shot, few-shot, and fine-tuning) applied to the DeepSeek, LLaMA 3.2, and Qwen 2.5 models. This ensured that the differences observed in the results could be attributed to the model's behavior and not to external factors such as training configuration or input/output size.

Table \ref{tab:binary_hyperparameters} shows the hyperparameters used in the binary classification task for the 3 LLMs. Since the expected response was a binary value (Attack or Normal), smaller values for $max\_new\_tokens$ and fewer training steps were used during fine-tuning.

\begin{table}[ht]
\centering
\caption{Hyperparameters used for binary classification}
\label{tab:binary_hyperparameters}
\begin{tabular}{|p{3cm}|p{12cm}|}
\hline
\textbf{Strategy} & \textbf{Hyperparameters and configuration} \\
\hline
\textbf{Zero-shot} &
\texttt{max\_seq\_length} = 2048, \texttt{dtype} = None, \texttt{load\_in\_4bit} = True. Model loaded in inference mode with Unsloth. \\
\hline
\textbf{Few-shot} &
Same loading parameters as in zero-shot. Generation with \texttt{max\_new\_tokens} = 3, using \texttt{use\_cache = True}. Decoding with \texttt{batch\_decode}. \\
\hline
\textbf{Fine-tuning} &
\texttt{max\_seq\_length} = 2048, \texttt{load\_in\_4bit} = True
\vspace{4pt}

\texttt{\textbf{LoRA config}: r = 16}, \texttt{lora\_alpha = 16}, \texttt{lora\_dropout = 0}, \texttt{target\_modules} = [Attention projections and FFN layers]

\vspace{4pt}
\texttt{\textbf{TrainingArguments}: }\texttt{batch\_size} = 2, \texttt{gradient\_accumulation\_steps} = 4, \texttt{learning\_rate} = 2e-4, \texttt{max\_steps} = 500, \texttt{warmup\_steps} = 5, \texttt{weight\_decay} = 0.01, \texttt{fp16/bf16} depending on availability, \texttt{optim} = adamw\_8bit. Evaluation strategies and saving every N steps, with \texttt{early stopping} after 3 evaluations without improvement.
\\
\hline
\end{tabular}
\end{table}

Table \ref{tab:multiclass_hyperparameters} summarizes the hyperparameters used in the multi-class classification task. Since the model had to identify one of the fourteen types of attacks or a normal one, a longer input sequence was used in few-shot ($max\_seq\_length$) and a higher number of steps during fine-tuning ($max\_steps$) to promote the learning of the class space.

\begin{table}[ht]
\centering
\caption{Hyperparameters used for multi-class classification}
\label{tab:multiclass_hyperparameters}
\begin{tabular}{|p{3cm}|p{12cm}|}
\hline
\textbf{Strategy} & \textbf{Hyperparameters and configuration} \\
\hline
\textbf{Zero-shot} &
\texttt{max\_seq\_length} = 2048, \texttt{dtype} = None, \texttt{load\_in\_4bit} = True. Model loaded with \texttt{FastLanguageModel.from\_pretrained} using Unsloth. \\
\hline
\textbf{Few-shot} &
Same loading parameters as in zero-shot.
Generation with \texttt{max\_new\_tokens} = 100 to allow responses with full attack class names. Using \texttt{use\_cache = True} and deconding with \texttt{batch\_decode}. \\
\hline
\textbf{Fine-tuning} &
\texttt{max\_seq\_length} = 2048, \texttt{load\_in\_4bit} = True

\vspace{4pt} 
\texttt{LoRA config}: \texttt{r = 16}, \texttt{lora\_alpha = 16}, \texttt{lora\_dropout = 0}, \texttt{target\_modules} = [Attention projections and FFN layers]

\vspace{4pt}
\texttt{\textbf{TrainingArguments}:}
\texttt{batch\_size} = 2, \texttt{gradient\_accumulation\_steps} = 4, 
\texttt{learning\_rate} = 2e-4, \texttt{max\_steps} = 1000, 
\texttt{warmup\_steps} = 5, \texttt{weight\_decay} = 0.01. 
\texttt{fp16/bf16} depending on availability, \texttt{optim} = adamw\_8bit. 
Evaluation strategies and saving every N step, with \texttt{early stopping} after 3 evaluations without improvement.

\\
\hline
\end{tabular}
\end{table}

\subsection{Adding mitigation actions to the models}
For the generation of mitigation actions, a mapping was created between the attacks in the Edge-IIoTset dataset and the attack patterns from the CAPEC framework as explained in previous sections. Based on this mapping, the attack identifiers and descriptions were combined with the general mitigations provided by CAPEC. After cleaning the descriptions, instructional prompts were structured, requesting a specific mitigation adapted to IoT environments.

The generation was carried out using the DeepSeek-R1-Distill-Llama-8B model\footnote{\url{https://huggingface.co/unsloth/DeepSeek-R1-Distill-Llama-8B}} in inference mode, loaded via the FastLanguageModel utility from Unsloth, with a maximum sequence length of 2048 tokens and 4-bit loading to optimize resource usage. The generation configuration employed $temperature = 0.7$, $top\_p = 0.9$, $do\_sample = True$, $max\_new\_tokens = 200$, and $repetition\_penalty = 1.1$, with the aim of obtaining varied but coherent responses.

Each prompt was constructed with a standardized format that included the attack type, its definition, and general mitigations from the CAPEC framework, as well as an execution context in IoT environments.

The generated responses were stored in a list, processed to eliminate formatting errors, style inconsistencies, and linguistic noise. These samples constituted the final set used for the supervised fine-tuning of the specific mitigation generation task.

The fine-tuning was performed using the three LLM models employed in previous tasks. In this case, the model not only had to identify the type of attack but also propose mitigations directly. To ensure consistency in the experiments, the same hyperparameters used in classification were maintained, although the value of $max\_new\_tokens$ was increased to allow the generation of multiple lines of output.

\section{Results}
\label{results}
\subsection{Binary classification}
All classic machine learning models, such as Random Forest and XGBoost, achieved perfect performance in the binary classification task, with values of 100\% in all evaluated metrics. This result suggests that the dataset has sufficiently discriminative features between classes for traditional models to achieve a clear separation.

In the zero-shot approach, the results were significantly lower compared to those of the machine learning models. As shown in Table \ref{tab:zero_shot_binary}, LLaMA achieved the best overall performance (F1 Score = 0.4028), followed by DeepSeek and Qwen, whose performance was clearly inferior (F1 Score = 0.2819). These results reflect the difficulty of the task when the model has not been adapted to the domain or exposed to similar examples beforehand.

\begin{table}[ht]
\centering
\caption{Binary classification results with LLMs}
\label{tab:zero_shot_binary}
\begin{tabular}{|l|c|c|c|c|}
\hline
\textbf{Model} & \textbf{Accuracy} & \textbf{F1 Score} & \textbf{Precision} & \textbf{Recall}  \\
\hline
\multicolumn{5}{|c|}{Zero-shot}\\
\hline
DeepSeek & 0.5100 & 0.3529 & 0.3620 & 0.5100  \\
LLaMA    & \textbf{0.5139} &\ textbf{0.4028} & \textbf{0.4991} & \textbf{0.5139}   \\
Qwen     & 0.3878 & 0.2819 & 0.2471 & 0.3878  \\
\hline
\multicolumn{5}{|c|}{Few-shot}\\
\hline
DeepSeek & 0.6567 & 0.6564 & 0.6565 & 0.6567 \\
LLaMA    & 0.4567 & 0.4542 & 0.4585 & 0.4567  \\
Qwen     & \textbf{0.9078}  & \textbf{0.9067} & \textbf{0.9214} & \textbf{0.9078}\\
\hline
\end{tabular}
\end{table}

In the few-shot scenario, the performance of the models improved significantly compared to zero-shot. Qwen clearly stood out with an F1 Score of 0.9067, demonstrating considerable ability to generalize from just a few examples, which is quite surprising. DeepSeek showed intermediate performance (F1 = 0.6564), while LLaMA achieved lower results, with metrics very close to random (F1 = 0.4542). These results confirm the sensitivity of LLMs to the number and quality of the examples provided.

Similar to the results from classic machine learning, the three LLM models evaluated through fine-tuning also achieved perfect performance in binary classification. DeepSeek, LLaMA, and Qwen achieved 100\% accuracy in all evaluated metrics, which shows that once the models are adapted to the specific domain, they are capable of learning the relationships in a deterministic manner in this case.

\subsection{Multi-class classification}
This section describes the performance of the models in the multi-class classification task. As shown in Table~\ref{tab:multiclass_zero_few}, zero-shot and few-shot strategies yield very limited results, with all models achieving F1-scores below 0.07. Although Qwen reached the highest accuracy in zero-shot mode (6.6\%), and DeepSeek performed best in few-shot (F1 of 0.0665), overall performance remained far from acceptable for real-world deployment.

\begin{table}[ht]
\centering
\caption{Multi-class classification results for LLMs in few and zero-shot modes.}
\label{tab:multiclass_zero_few}
\begin{tabular}{|p{1.4cm}|c|c|c|c|}
\hline
\textbf{Model} & \textbf{Accuracy} & \textbf{F1 Score} & \textbf{Precision} & \textbf{Recall} \\
\hline
\multicolumn{5}{|c|}{Zero-shot}\\
\hline
DeepSeek & 0.0418 & 0.0065 & 0.0035 & 0.0418 \\
LLaMA    & 0.0200 & \textbf{0.0126} & \textbf{0.0095} & 0.0200 \\
Qwen     & \textbf{0.0662} & 0.0120 & 0.0066 & \textbf{0.0662} \\
\hline
\multicolumn{5}{|c|}{Few-shot}\\
\hline
DeepSeek & \textbf{0.0782} & \textbf{0.0665} & 0.0868 & \textbf{0.0782} \\
LLaMA    & 0.0071 & 0.0117 & \textbf{0.1229} & 0.0071 \\
Qwen     & 0.0142 & 0.0178 & 0.0966 & 0.0142 \\
\hline
\end{tabular}
\end{table}

When fine-tuning is applied, the performance of LLMs improves dramatically compared to zero-shot and few-shot modes. Table~\ref{tab:multiclass_less} shows the results using fewer training samples. In this setting, DeepSeek clearly outperforms all other models with an F1-score of 0.7479 and precision above 80\%. LLaMA also delivers competitive results despite its smaller size, while Qwen lags behind the other LLMs, although still comparable to traditional ML models.

\begin{table}[ht]
\centering
\caption{Multi-class classification results for ML and LLMs fine-tuned models (\textbf{fewer samples}).}
\label{tab:multiclass_less}
\begin{tabular}{|p{1.4cm}|c|c|c|c|}
\hline
\textbf{Model} & \textbf{Accuracy} & \textbf{F1 Score} & \textbf{Precision} & \textbf{Recall} \\
\hline
\multicolumn{5}{|c|}{ML models}\\
\hline
MLP           & 0.2373 & 0.1774 & 0.2658 & 0.2373 \\
XGBoost       & \textbf{0.5164} & \textbf{0.4987} & 0.6325 & \textbf{0.5164} \\
LightGBM      & 0.5147 & 0.4942 & \textbf{0.6335} & 0.5147 \\
R. Forest  & 0.4907 & 0.4554 & 0.5913 & 0.4907 \\
\hline
\multicolumn{5}{|c|}{LLMs fine-tuned}\\
\hline
DeepSeek & \textbf{0.7502} & \textbf{0.7479} & \textbf{0.8009} & \textbf{0.7502} \\
LLaMA    & 0.6942 & 0.6789 & 0.6776 & 0.6942 \\
Qwen     & 0.4756 & 0.4546 & 0.5974 & 0.4756 \\
\hline
\end{tabular}
\end{table}

In the larger dataset scenario (Table~\ref{tab:multiclass_more}), DeepSeek maintains its lead with an F1-score of 0.7154, followed closely by LLaMA. Interestingly, DeepSeek also performs well when trained on fewer samples, suggesting a strong ability to generalize under data-constrained conditions. The performance of the ML models shows minimal change between both scenarios, with XGBoost and LightGBM remaining the top classical approaches.

\begin{table}[ht]
\centering
\caption{Multi-class classification results for ML and LLMs fine-tuned models (\textbf{more samples}).}
\label{tab:multiclass_more}
\begin{tabular}{|p{1.4cm}|c|c|c|c|}
\hline
\textbf{Model} & \textbf{Accuracy} & \textbf{F1 Score} & \textbf{Precision} & \textbf{Recall} \\
\hline
\multicolumn{5}{|c|}{ML models}\\
\hline
MLP           & 0.2573 & 0.2018 & 0.3883 & 0.2573 \\
XGBoost       & \textbf{0.5191} & \textbf{0.4937} & \textbf{0.6314} & \textbf{0.5191} \\
LightGBM      & 0.5160 & 0.4933 & 0.6188 & 0.5160 \\
R. Forest  & 0.4787 & 0.4388 & 0.6286 & 0.4787 \\
\hline
\multicolumn{5}{|c|}{LLMs fine-tuned}\\
\hline
DeepSeek & \textbf{0.7320} & \textbf{0.7154} & \textbf{0.7573} & \textbf{0.7320} \\
LLaMA    & 0.7036 & 0.7048 & 0.7307 & 0.7036 \\
Qwen     & 0.5867 & 0.5940 & 0.6911 & 0.5867 \\
\hline
\end{tabular}
\end{table}

In summary, fine-tuned LLMs consistently outperform both their zero/few-shot counterparts and classical ML models in multi-class classification. DeepSeek demonstrates the best overall performance, even when trained on fewer samples, highlighting its strong generalization capabilities. LLaMA also offers competitive results given its smaller size, making it suitable for constrained environments. In contrast, traditional ML models show limited improvement with increased data, confirming the advantage of language models when properly fine-tuned for IoT security tasks.

\subsection{Detailed Comparison: DeepSeek vs. XGBoost}

To better assess the effectiveness of LLMs versus traditional models, we compared the best-performing classical model (XGBoost) with the best LLM (DeepSeek) under two evaluation settings: imbalanced and balanced subsets of the test data.

\paragraph{Imbalanced inference} 
As shown in Table~\ref{tab:inferencia_desbalanceadas}, DeepSeek achieved the highest overall performance, even when trained with fewer samples. While XGBoost also performed well—likely due to the overrepresentation of the \textit{Normal} class—DeepSeek demonstrated higher F1-score and recall, highlighting its better generalization in multiclass scenarios. However, detailed per-class analysis (Table~\ref{tab:deepseek_per_class}) revealed that DeepSeek struggled with rare classes such as \textit{MiTM} and \textit{Fingerprinting}, which were nearly undetected.

\begin{table}[ht]
\centering
\caption{Inference results with randomly sampled imbalanced data (21,000 samples)}
\label{tab:inferencia_desbalanceadas}
\begin{tabular}{|p{1.25cm}|c|c|c|c|c|}
\hline
\textbf{Model} & \textbf{Train. samples} & \textbf{Accuracy} & \textbf{F1 Score} & \textbf{Precision} & \textbf{Recall}\\
\hline
XGBoost    & 10.5k & 0.8667 & 0.8472 & 0.8553 & 0.8667 \\
XGBoost    & 5.25k & 0.8656 & 0.8470 & 0.8534 & 0.8656 \\
DeepSeek   & 10.5k & 0.9051 & 0.9122 & 0.9325 & 0.9051 \\
DeepSeek   & 5.25k & \textbf{0.9255} & \textbf{0.9250} & \textbf{0.9347} & \textbf{0.9255} \\
\hline
\end{tabular}
\end{table}

\begin{table}[ht]
\centering
\caption{Per-class performance of the DeepSeek model during inference (imbalanced samples)}
\label{tab:deepseek_per_class}
\begin{tabular}{|l|c|c|c|}
\hline
\textbf{Attack class} & \textbf{F1 Score} & \textbf{Precision} & \textbf{Recall} \\
\hline
Normal                    & \textbf{0.9993} & 1.0000 & 0.9986 \\
DDoS\_UDP                 & 0.5823          & 0.5058 & 0.6860 \\
DDoS\_ICMP                & 0.3015          & 0.5381 & 0.2095 \\
SQL\_injection            & 0.9912          & 0.9883 & 0.9941 \\
Password                 & 0.9907          & 1.0000 & 0.9815 \\
Vulnerability\_scanner    & 0.9947          & 0.9936 & 0.9957 \\
DDoS\_TCP                 & 0.7975          & 0.7594 & 0.8396 \\
DDoS\_HTTP                & 0.9882          & 0.9863 & 0.9902 \\
Uploading                & 0.9193          & 0.9608 & 0.8812 \\
Backdoor                 & 0.8660          & 0.7749 & 0.9813 \\
Port\_Scanning            & 0.3395          & 0.4955 & 0.2582 \\
XSS                      & 0.9225          & 0.9690 & 0.8803 \\
Ransomware               & 0.8218          & 0.8300 & 0.8137 \\
MiTM                     & 0.0000          & 0.0000 & 0.0000 \\
Fingerprinting           & 0.0217          & 0.0112 & 0.3636 \\
\hline
\end{tabular}
\end{table}

\paragraph{Balanced inference}
When evaluated on a balanced subset (Table~\ref{tab:inferencia_balanceadas}), DeepSeek clearly outperformed XGBoost across all metrics. Notably, the model trained on fewer samples slightly surpassed the one trained with more, indicating better generalization. Per-class metrics (Table~\ref{tab:deepseek_per_class_balanced}) confirm excellent performance on frequent classes like \textit{Password}, \textit{SQL\_injection}, and \textit{Vulnerability\_scanner}, while rare or noisy classes such as \textit{MiTM} and \textit{DDoS\_ICMP} remain challenging. 

\begin{table}[ht]
\centering
\caption{Inference results with randomly sampled balanced data (20,415 samples)}
\label{tab:inferencia_balanceadas}
\begin{tabular}{|p{1.25cm}|c|c|c|c|c|}
\hline
\textbf{Model} & \textbf{Train. samples} & \textbf{Accur.} & \textbf{F1 Score} & \textbf{Prec.} & \textbf{Recall}\\
\hline
XGBoost    & 10.5k    & 0.5383 & 0.5147 & 0.6444 & 0.5383 \\
XGBoost    & 5.25k  & 0.5372 & 0.5169 & 0.6362 & 0.5372 \\
DeepSeek   & 10.5k    & 0.7404 & 0.7239 & 0.7496 & 0.7404 \\
DeepSeek   & 5.25k  & \textbf{0.7487} & \textbf{0.7423} & \textbf{0.7852} & \textbf{0.7487} \\
\hline
\end{tabular}
\end{table}

\begin{table}[ht]
\centering
\caption{Per-class performance of the DeepSeek model during inference (balanced samples)}
\label{tab:deepseek_per_class_balanced}
\begin{tabular}{|l|c|c|c|}
\hline
\textbf{Attack class} & \textbf{F1 Score} & \textbf{Precision} & \textbf{Recall} \\
\hline
Normal                    & \textbf{0.9982} & 1.0000 & 0.9964 \\
DDoS\_UDP                 & 0.4212          & 0.3100 & 0.6571 \\
DDoS\_ICMP                & 0.2355          & 0.3153 & 0.1879 \\
SQL\_injection            & 0.9784          & 0.9692 & 0.9879 \\
Password                 & 0.9856          & 0.9956 & 0.9757 \\
Vulnerability\_scanner    & 0.9961          & 1.0000 & 0.9921 \\
DDoS\_TCP                 & 0.6604          & 0.5511 & 0.8236 \\
DDoS\_HTTP                & 0.9774          & 0.9652 & 0.9900 \\
Uploading                & 0.9104          & 0.9531 & 0.8714 \\
Backdoor                 & 0.8668          & 0.7857 & 0.9664 \\
Port\_Scanning            & 0.3603          & 0.6446 & 0.2500 \\
XSS                      & 0.9349          & 0.9819 & 0.8921 \\
Ransomware               & 0.8915          & 0.9221 & 0.8629 \\
MiTM                     & 0.4844          & 1.0000 & 0.3196 \\
Fingerprinting           & 0.2626          & 0.2644 & 0.2607 \\
\hline
\end{tabular}
\end{table}

%\begin{figure}[ht]
%    \centering
%    \includegraphics[width=8cm]{images/confusion_matrix_llama.png}
%    \caption{Confusion matrix of the fine-tuned Llama model for multi-class}
%    \label{fig:cm_llama}
%\end{figure}

%\begin{figure}[ht]
%    \centering
%    \includegraphics[width=8cm]{images/confusion_matrix_qwen.png}
%    \caption{Confusion matrix of the fine-tuned Qwen model for multi-class}
%    \label{fig:cm_qwen}
%\end{figure}

\subsection{Detection with mitigation actions}

The mitigation generation results are especially positive for the DeepSeek and LLaMA models, which achieved 100\% quality in multiple classes according to ROUGE-L and Cosine Similarity metrics. Notably, these models often generated correct mitigations even when the predicted attack type was incorrect, demonstrating strong semantic generalization.

In contrast, Qwen did not reach the same level of performance. One plausible explanation is that the initial mitigation samples used for fine-tuning were generated using DeepSeek, which is based on the LLaMA architecture, potentially favoring both during training. As shown in Table~\ref{tab:qwen_mitigations_quality}, Qwen still obtains high scores in both metrics, but does not consistently reproduce the expected mitigations with perfect alignment.

\begin{table}[ht]
\centering
\caption{Quality of mitigations generated by Qwen per attack class}
\label{tab:qwen_mitigations_quality}
\begin{tabular}{|l|c|c|}
\hline
\textbf{Attack class} & \textbf{ROUGE-L} & \textbf{Cosine Similarity} \\
\hline
Backdoor               & 0.9957 & 0.9997 \\
Vulnerability\_scanner & 1.0000 & 1.0000 \\
SQL\_injection         & 0.9853 & 0.9995 \\
DDoS\_UDP              & 1.0000 & 1.0000 \\
Port\_Scanning         & 0.9894 & 0.9992 \\
DDoS\_HTTP             & 1.0000 & 1.0000 \\
DDoS\_TCP              & 1.0000 & 1.0000 \\
MITM                   & 1.0000 & 1.0000 \\
DDoS\_ICMP             & 1.0000 & 1.0000 \\
Ransomware             & 1.0000 & 1.0000 \\
Password               & 0.8862 & 0.9582 \\
Fingerprinting         & 1.0000 & 1.0000 \\
XSS                    & 1.0000 & 1.0000 \\
Uploading              & 1.0000 & 1.0000 \\
\hline
\end{tabular}
\end{table}

\section{Conclusions}
\label{conclusions}
In this paper, we have explored the application of large language models (LLMs) to anomaly detection and mitigation generation in IoT security logs. Using the Edge-IIoTset dataset, we conducted a complete methodology that includes data preprocessing, feature selection, and the evaluation of multiple models. Two classification tasks were addressed: binary detection (malicious vs. benign) and multi-class classification (attack type identification). The dataset was normalized and structured for LLM input, and several model variants were trained and assessed.

We compared the performance of classical machine learning models (Random Forest, XGBoost, LightGBM, MLP) with three open-source LLMs (DeepSeek-R1, LLaMA 3.2, Qwen 2.5) under zero-shot, few-shot, and fine-tuning configurations. While both model families reached perfect scores in binary classification when sufficient training data was available, LLMs significantly outperformed classical models in multi-class tasks. DeepSeek and LLaMA consistently ranked highest, with DeepSeek reaching up to 0.7479 F1-score even when trained on fewer samples.

We also designed a mitigation generation module using the MITRE CAPEC framework. DeepSeek was used to generate IoT-adapted mitigation actions based on general CAPEC attack descriptions. These specific samples were then used to fine-tune the three LLMs. The models, especially DeepSeek and LLaMA, achieved perfect or near-perfect performance in semantic and lexical evaluation metrics (ROUGE-L and cosine similarity), confirming their ability to not only detect threats but also produce context-aware responses—an aspect rarely addressed in previous work.

DeepSeek was the most robust and consistent model across all tasks, with perfect or near-perfect scores in both classification and mitigation generation. LLaMA also performed remarkably well despite its smaller size, making it attractive for deployment in constrained environments such as edge or embedded systems.

Overall, our results demonstrate that fine-tuned LLMs offer a highly effective solution for both detection and response in IoT security logs. They outperform traditional approaches in multi-class classification and introduce the added value of generating actionable mitigations.

\section*{Acknowledgments}
The authors would like to acknowledge the support of R\&D project PID2022-136684OB-C21 (Fun4Date) funded by the Spanish Ministry of Science and Innovation MCIN/AEI/ 10.13039/501100011033 and TUCAN6-CM (TEC-2024/COM-460), funded by CM (ORDEN 5696/2024)

\bibliographystyle{IEEEtran}

\bibliography{references}

\begin{thebibliography}{10}
\providecommand{\url}[1]{#1}
\csname url@samestyle\endcsname
\providecommand{\newblock}{\relax}
\providecommand{\bibinfo}[2]{#2}
\providecommand{\BIBentrySTDinterwordspacing}{\spaceskip=0pt\relax}
\providecommand{\BIBentryALTinterwordstretchfactor}{4}
\providecommand{\BIBentryALTinterwordspacing}{\spaceskip=\fontdimen2\font plus
\BIBentryALTinterwordstretchfactor\fontdimen3\font minus \fontdimen4\font\relax}
\providecommand{\BIBforeignlanguage}[2]{{%
\expandafter\ifx\csname l@#1\endcsname\relax
\typeout{** WARNING: IEEEtran.bst: No hyphenation pattern has been}%
\typeout{** loaded for the language `#1'. Using the pattern for}%
\typeout{** the default language instead.}%
\else
\language=\csname l@#1\endcsname
\fi
#2}}
\providecommand{\BIBdecl}{\relax}
\BIBdecl

\bibitem{xu_detecting_2009}
W.~Xu, L.~Huang, A.~Fox, D.~Patterson, and M.~I. Jordan, ``Detecting large-scale system problems by mining console logs,'' in \emph{Proceedings of the ACM SIGOPS 22nd Symposium on Operating Systems Principles}, 2009, pp. 117--132.

\bibitem{liu_isolation_2008}
F.~T. Liu, K.~M. Ting, and Z.-H. Zhou, ``Isolation forest,'' in \emph{2008 Eighth IEEE International Conference on Data Mining}.\hskip 1em plus 0.5em minus 0.4em\relax IEEE, 2008, pp. 413--422.

\bibitem{wang_anomaly_2004}
Y.~Wang, J.~Wong, and A.~Miner, ``Anomaly intrusion detection using one class svm,'' in \emph{Proceedings from the Fifth Annual IEEE SMC Information Assurance Workshop}.\hskip 1em plus 0.5em minus 0.4em\relax IEEE, 2004, pp. 358--364.

\bibitem{han_loggpt_2023}
X.~Han, S.~Yuan, and M.~Trabelsi, ``{LogGPT}: Log anomaly detection via {GPT},'' in \emph{2023 {IEEE} International Conference on Big Data ({BigData})}, 2023, pp. 1117--1122.

\bibitem{cui_logeval_2024}
\BIBentryALTinterwordspacing
T.~Cui, S.~Ma, Z.~Chen, T.~Xiao, S.~Tao, Y.~Liu, S.~Zhang, D.~Lin, C.~Liu, Y.~Cai, W.~Meng, Y.~Sun, and D.~Pei, ``{LogEval}: A comprehensive benchmark suite for large language models in log analysis,'' \_eprint: 2407.01896. [Online]. Available: \url{https://arxiv.org/abs/2407.01896}
\BIBentrySTDinterwordspacing

\bibitem{akhtar_llm-based_2025}
\BIBentryALTinterwordspacing
S.~Akhtar, S.~Khan, and S.~Parkinson, ``{LLM}-based event log analysis techniques: A survey,'' \_eprint: 2502.00677. [Online]. Available: \url{https://arxiv.org/abs/2502.00677}
\BIBentrySTDinterwordspacing

\bibitem{ferrag_revolutionizing_2024}
M.~A. Ferrag, M.~Ndhlovu, N.~Tihanyi, L.~C. Cordeiro, M.~Debbah, T.~Lestable, and N.~S. Thandi, ``Revolutionizing cyber threat detection with large language models: A privacy-preserving {BERT}-based lightweight model for {IoT}/{IIoT} devices,'' \emph{{IEEE} Access}, vol.~12, pp. 23\,733--23\,750, 2024.

\bibitem{toniot}
A.~A. et~al., ``Ton\_iot datasets,'' \url{https://research.unsw.edu.au/projects/toniot-datasets}, 2020, accessed: 2025-04-16.

\bibitem{botiot}
\BIBentryALTinterwordspacing
N.~Koroniotis, N.~Moustafa, E.~Sitnikova, and B.~Turnbull, ``Towards the development of realistic botnet dataset in the internet of things for network forensic analytics: Bot-iot dataset,'' \emph{Future Generation Computer Systems}, vol. 100, pp. 779--796, 2019. [Online]. Available: \url{https://www.sciencedirect.com/science/article/pii/S0167739X18327687}
\BIBentrySTDinterwordspacing

\bibitem{tian_cldtlog_2023}
\BIBentryALTinterwordspacing
G.~Tian, N.~Luktarhan, H.~Wu, and Z.~Shi, ``{CLDTLog}: System log anomaly detection method based on contrastive learning and dual objective tasks,'' \emph{Sensors}, vol.~23, no.~11, 2023. [Online]. Available: \url{https://www.mdpi.com/1424-8220/23/11/5042}
\BIBentrySTDinterwordspacing

\bibitem{guo_logbert_2021}
H.~Guo, S.~Yuan, and X.~Wu, ``{LogBERT}: Log anomaly detection via {BERT},'' in \emph{2021 International Joint Conference on Neural Networks ({IJCNN})}, 2021, pp. 1--8.

\bibitem{du_deeplog_2017}
\BIBentryALTinterwordspacing
M.~Du, F.~Li, G.~Zheng, and V.~Srikumar, ``Deeplog: Anomaly detection and diagnosis from system logs through deep learning,'' in \emph{Proceedings of the 2017 ACM SIGSAC Conference on Computer and Communications Security (CCS '17)}.\hskip 1em plus 0.5em minus 0.4em\relax Association for Computing Machinery, 2017, pp. 1285--1298. [Online]. Available: \url{https://doi.org/10.1145/3133956.3134015}
\BIBentrySTDinterwordspacing

\bibitem{zhou_loganomaly_2019}
R.~Zhou, P.~Sun, S.~Tao, R.~Zhang, W.~Meng, Y.~Liu, Y.~Zhu, Y.~Liu, D.~Pei, S.~Zhang, and Y.~Chen, ``Loganomaly: Unsupervised detection of sequential and quantitative anomalies in unstructured logs,'' in \emph{Proceedings of the 28th International Joint Conference on Artificial Intelligence (IJCAI)}, 2019, pp. 4739--4745.

\bibitem{oliner_what_2007}
A.~Oliner and J.~Stearley, ``What supercomputers say: A study of five system logs,'' in \emph{37th Annual IEEE/IFIP International Conference on Dependable Systems and Networks (DSN '07)}.\hskip 1em plus 0.5em minus 0.4em\relax IEEE, 2007, pp. 575--584.

\bibitem{hadadi_llm_2025}
\BIBentryALTinterwordspacing
F.~Hadadi, Q.~Xu, D.~Bianculli, and L.~Briand, ``{LLM} meets {ML}: Data-efficient anomaly detection on unseen unstable logs,'' \_eprint: 2406.07467. [Online]. Available: \url{https://arxiv.org/abs/2406.07467}
\BIBentrySTDinterwordspacing

\bibitem{balanfine}
G.~Balan, C.-A. Simion, and D.~T. Gavrilut, ``Fine-tuning llms vs non-generative machine learning models: A comparative study of malware detection,'' in \emph{Proceedings of the 17th International Conference on Agents and Artificial Intelligence (ICAART)}.\hskip 1em plus 0.5em minus 0.4em\relax ScitePress, 2025.

\bibitem{iot23}
S.~G. et~al., ``Iot-23 dataset,'' \url{https://www.stratosphereips.org/datasets-iot23}, 2020, accessed: 2025-04-16.

\bibitem{cicdatasets}
C.~I. for Cybersecurity, ``Cic datasets,'' \url{https://www.unb.ca/cic/datasets/index.html}, 2021, accessed: 2025-04-16.

\bibitem{rtiot2022}
``Rt-iot 2022: Real-time internet of things dataset,'' \url{https://archive.ics.uci.edu/dataset/942/rt-iot2022}, accessed: 2025-04-16.

\bibitem{edgeiiot}
M.~A.~F. et~al., ``Edge-iiotset: Cyber security dataset of iot \& iiot,'' \url{https://www.kaggle.com/datasets/mohamedamineferrag/edgeiiotset-cyber-security-dataset-of-iot-iiot}, accessed: 2025-04-16.

\bibitem{guastalla_application_2024}
M.~Guastalla, Y.~Li, A.~Hekmati, and B.~Krishnamachari, ``Application of large language models to {DDoS} attack detection,'' in \emph{Security and Privacy in Cyber-Physical Systems and Smart Vehicles}, Y.~Chen, C.-W. Lin, B.~Chen, and Q.~Zhu, Eds.\hskip 1em plus 0.5em minus 0.4em\relax Springer Nature Switzerland, 2024, pp. 83--99.

\bibitem{zhang_hackmentor_2023}
J.~Zhang, H.~Wen, L.~Deng, M.~Xin, Z.~Li, L.~Li, H.~Zhu, and L.~Sun, ``{HackMentor}: Fine-tuning large language models for cybersecurity,'' in \emph{2023 {IEEE} 22nd International Conference on Trust, Security and Privacy in Computing and Communications ({TrustCom})}, 2023, pp. 452--461.

\bibitem{hu2021loralowrankadaptationlarge}
\BIBentryALTinterwordspacing
E.~J. Hu, Y.~Shen, P.~Wallis, Z.~Allen-Zhu, Y.~Li, S.~Wang, L.~Wang, and W.~Chen, ``Lora: Low-rank adaptation of large language models,'' 2021. [Online]. Available: \url{https://arxiv.org/abs/2106.09685}
\BIBentrySTDinterwordspacing

\bibitem{zhang_llm4itd_2024}
M.~Zhang, X.~Liang, F.~Tian, Y.~Yang, H.~Yu, and B.~Li, ``{LLM}4itd: Insider threat detection with fine-tuned large language models,'' in \emph{2024 International Conference on Interactive Intelligent Systems and Techniques ({IIST})}, 2024, pp. 236--241.

\bibitem{batmaz_building_2024}
\BIBentryALTinterwordspacing
G.~Batmaz, ``Building cyber language models to unlock new cybersecurity capabilities.'' [Online]. Available: \url{https://developer.nvidia.com/blog/building-cyber-language-models-to-unlock-new-cybersecurity-capabilities/}
\BIBentrySTDinterwordspacing

\bibitem{karlsen_benchmarking_2024}
\BIBentryALTinterwordspacing
E.~Karlsen, X.~Luo, N.~Zincir-Heywood, and M.~Heywood, ``Benchmarking large language models for log analysis, security, and interpretation,'' \emph{Journal of Network and Systems Management}, vol.~32, no.~3, p.~59, 2024. [Online]. Available: \url{https://doi.org/10.1007/s10922-024-09831-x}
\BIBentrySTDinterwordspacing

\bibitem{labs_building_2025}
\BIBentryALTinterwordspacing
I.~Labs, ``Building a dynamic parserless network log and security alert platform with {LLM} and pydantic.'' [Online]. Available: \url{https://ithy.com/article/llm-pydantic-log-parser-security-kia8dv2t}
\BIBentrySTDinterwordspacing

\bibitem{yang_security_2024}
\BIBentryALTinterwordspacing
A.~Z.~H. Yang, H.~Tian, H.~Ye, R.~Martins, and C.~L. Goues, ``Security vulnerability detection with multitask self-instructed fine-tuning of large language models,'' \_eprint: 2406.05892. [Online]. Available: \url{https://arxiv.org/abs/2406.05892}
\BIBentrySTDinterwordspacing

\bibitem{li_redchronos_2025}
\BIBentryALTinterwordspacing
C.~Li, Z.~Zhu, J.~He, and X.~Zhang, ``{RedChronos}: A large language model-based log analysis system for insider threat detection in enterprises,'' \_eprint: 2503.02702. [Online]. Available: \url{https://arxiv.org/abs/2503.02702}
\BIBentrySTDinterwordspacing

\bibitem{boffa_logprecis_2024}
\BIBentryALTinterwordspacing
M.~Boffa, I.~Drago, M.~Mellia, L.~Vassio, D.~Giordano, R.~Valentim, and Z.~B. Houidi, ``{LogPrécis}: Unleashing language models for automated malicious log analysis: Précis: A concise summary of essential points, statements, or facts,'' \emph{Computers \& Security}, vol. 141, p. 103805, 2024. [Online]. Available: \url{https://www.sciencedirect.com/science/article/pii/S0167404824001068}
\BIBentrySTDinterwordspacing

\bibitem{obrien_mitre_2024}
T.~O'Brien, ``Mitre att\&ck labeling of cyber threat intelligence via llm,'' \url{https://sansorg.egnyte.com/dl/I7dMyUXrb3}, 2024, sANS Institute, diciembre 2024.

\bibitem{zhang_when_2025}
\BIBentryALTinterwordspacing
Y.~Zhang, Z.~Chen, X.~Liu, B.~Zhang, L.~Wang, and J.~Wu, ``When llms meet cybersecurity: A systematic literature review,'' \emph{Cybersecurity}, vol.~8, p.~55, 2025. [Online]. Available: \url{https://cybersecurity.springeropen.com/articles/10.1186/s42400-025-00361-w}
\BIBentrySTDinterwordspacing

\bibitem{mitre_capec}
\BIBentryALTinterwordspacing
{MITRE}, ``Capec: Common attack pattern enumeration and classification,'' Último acceso: abril de 2024. [Online]. Available: \url{https://capec.mitre.org/}
\BIBentrySTDinterwordspacing

\bibitem{huggingface2023}
H.~Face, ``Transformers: State-of-the-art machine learning for pytorch, tensorflow, and jax,'' 2023, \url{https://huggingface.co}.

\bibitem{unsloth2024}
{Unsloth Team}, ``Unsloth,'' 2024, \url{https://unsloth.ai/}.

\bibitem{deepseek2024}
{DeepSeek AI}, ``Deepseek,'' 2024, \url{https://huggingface.co/deepseek-ai}.

\bibitem{llama32024}
{Meta AI}, ``Llama,'' 2024, \url{https://huggingface.co/meta-llama}.

\bibitem{qwen2024}
{Alibaba DAMO Academy}, ``Qwen,'' 2024, \url{https://huggingface.co/Qwen}.

\end{thebibliography}

\end{document}